\documentclass[aps,prd,preprint,floatfix,nofootinbib]{revtex4}
\usepackage{graphicx}
\usepackage{color}

\begin{document}



\title{
Hidden fermion as milli-charged dark matter \\
in Stueckelberg $Z'$ model}

\renewcommand{\thefootnote}{\fnsymbol{footnote}}

\author{ 
Kingman Cheung
\footnote{Email address: cheung@phys.nthu.edu.tw}
and Tzu-Chiang Yuan
\footnote{Email address: tcyuan@phys.nthu.edu.tw} }
\affiliation{Department of Physics, National Tsing Hua University, 
Hsinchu, Taiwan \\
Physics Division, National Center for Theoretical Sciences, Hsinchu, Taiwan}

\renewcommand{\thefootnote}{\arabic{footnote}}
\date{\today}

\begin{abstract}
We augment the hidden Stueckelberg $Z'$ model by a pair of Dirac
fermions in the hidden sector, in which the $Z'$ has a coupling
strength comparable to weak scale coupling.  We show that this hidden
fermion-antifermion pair could be a milli-charged dark matter
candidate with a viable relic density.
Existing terrestrial and astrophysical searches on milli-charged particles 
do not place severe constraints on this hidden fermion. 
We calculate the flux of monochromatic photons coming from the
Galactic center due to pair annihilation of these milli-charged particles 
and show that it is within reach of the next generation
$\gamma$-ray experiments. 
The characteristic signature of this theoretical endeavor is that the 
Stueckelberg $Z'$ boson has a large invisible width decaying into the hidden 
fermion-antifermion pair. 
We show that existing Drell-Yan data do not constrain this model yet. 
Various channels of singly production of this $Z'$ boson
at the LHC and ILC are explored.
\end{abstract}

\maketitle
 
\section{Introduction}
The standard model (SM) of particle physics has been blessed with her
elegant way of giving masses to the weak gauge bosons by the Higgs
mechanism.  However, the crucial ingredient of this mechanism, the
Higgs boson, is still missing.  In addition, a scalar Higgs boson mass
is not stable under perturbative calculation.  It will receive an
enormous amount of radiative corrections to its mass such that a
delicate cancellation between its bare mass and radiative corrections
is needed so as to obtain a mass in the electroweak scale -- this is
the famous hierarchy problem.  An alternative way to give mass to an
abelian $U(1)$ gauge boson is known as the Stueckelberg mechanism.
Although it is very difficult to give masses to nonabelian gauge
bosons without sacrificing renormalizability within the Stueckelberg
approach, it is worthwhile to study the consequence of this mechanism
as an extension to the SM with extra abelian $U(1)$ factors.

Recently, Kors and Nath \cite{nath} showed that the SM extended by a
hidden sector described by a Stueckelberg $U(1)_X$ and the gauge field
$C_\mu$ associated with it can pass all the existing constraints from
electroweak data as well as direct search limits from the Tevatron.
Through the combined Stueckelberg and Higgs mechanisms, the SM
$SU_L(2) \times U(1)_Y$ gauge fields $B_\mu$ and $W^3_\mu$ mix with
the hidden sector gauge field $C_\mu$.  After rotation from the
interaction basis $(C_\mu, \, B_\mu,\, W^3_\mu)$ to the mass
eigenbasis $(Z'_\mu, Z_\mu, A_\mu)$, one obtains a massless state
identified to be the photon $\gamma$ and two massive eigenstates which
are the SM-like $Z$ boson and an additional $Z'$ boson.  As long as
the mixing is small, the $Z'$ boson only couples very weakly to the SM
fermions, and so it can evade all the existing constraints on
conventional $Z'$ models.  The allowed mass range for the Stueckelberg
$Z'$ can be anywhere from 200 GeV to a few TeV \cite{liu}.  Typically,
the mass of the Stueckelberg $Z'$ is above the SM $Z$ boson mass.

In this work, a pair of hidden Dirac fermions is introduced in the
Stueckelberg $Z'$ model.  Such a possibility has been mentioned in
Ref.\cite{nath}, but its phenomenology was not explored.  There could
be various types or generations of fermions in the hidden sector, just
like our visible world. Since the abelian $U(1)_X$ is assumed to be
the only gauge group in the hidden sector and there is no connector
sector between our visible world and the hidden one in this class of
models, all hidden fermions in this sector are stable.
\footnote
{
This is in analogous to the pure QED case, muon does not decay 
into an electron plus a photon. 
}
Thus the hidden fermion-antifermion pair that we add in the model 
can be viewed as the lightest ones in the 
hidden sector, should there be more than one type of them.
The SM fermions are neutral under this hidden $U(1)_X$.  
Since this hidden fermion pair is stable, it can be the dark matter candidate
of our Universe.  

In the next Section, we will present some details of the Stueckelberg
$Z'$ extension of the SM with an additional pair of Dirac
fermion-antifermion in the hidden sector.  In Section III, we discuss
milli-charged dark matter. Treating the hidden fermion as our
candidate of dark matter, we calculate its relic density and explore
the parameter space allowed by the WMAP measurement. We also calculate
the monochromatic photon flux coming from the Galactic center due to
pair annihilation of these hidden fermions.  In Section IV, we explore
some novel collider phenomenology of the Stueckelberg $Z'$ with the
presence of the hidden fermion. Since the width of the Stueckelberg
$Z'$ is no longer narrow, compared to the scenario studied in
\cite{liu}, its phenomenology is rather different.  Comments and
conclusions are given in Section V.

\section{The Model}
The Stueckelberg extension \cite{nath} of the SM (StSM) is obtained by adding a
hidden sector associated with an extra $U(1)_X$ interaction, under which the
SM particles are neutral.
\footnote
{It was shown in Ref. \cite{nath} that the SM fermions are neutral under
the extra $U(1)_X$ has the advantage of maintaining the neutron 
charge to be zero.
}
We explicitly specify the content of the hidden sector:
a gauge boson $C_\mu$ and a pair of fermion and antifermion
 $\chi$ and $\bar \chi$. 

The Lagrangian describing the system is ${\cal L}_{\rm StSM} ={\cal L}_{\rm SM} + 
{\cal L}_{\rm St}$, where
\begin{eqnarray}
{\cal L}_{\rm SM} &=& -\; \frac{1}{4} W_{\mu\nu}^a \, W^{a\mu\nu} 
                      - \frac{1}{4} B_{\mu\nu}\, B^{\mu\nu}
                   + D_\mu \Phi^\dagger \, D^\mu \Phi - V(\Phi^\dagger \,\Phi) 
   \nonumber \\
           && + \; i \bar \psi_f \gamma^\mu D_\mu \psi_f  \; ,\\
{\cal L}_{\rm St} &=& 
         - \frac{1}{4} C_{\mu\nu}\, C^{\mu\nu}
           + i \bar \chi \gamma^\mu D^X_\mu \chi  
 + \frac{1}{2} \left( \partial_\mu \sigma + M_1 C_\mu + M_2 B_\mu \right )^2
 \; ,
\\
 D_\mu &=& \partial_\mu + i g_2 \frac{\tau^a}{2} \,W^a_\mu + i g_Y \frac{Y}{2}
            B_\mu \label{dmu} \; ,\\
 D^X_\mu &=& \partial_\mu + i g_X \, Q_X \, C_\mu  \; ,
\label{dmuchi}
\end{eqnarray}
where $W_{\mu\nu}^a(a=1,2,3)$, $B_{\mu\nu}$, and $C_{\mu\nu}$ are the
field strength tensors of the gauge fields $W_\mu^a$, $B_\mu$, and
$C_\mu$, respectively.  The SM fermions $f$ were explicitly forbidden from
carrying the $U(1)_X$ charges, as implied by Eq. (\ref{dmu}), while the
hidden fermion pair only carries the $U(1)_X$ charge, as implied by
Eq. (\ref{dmuchi}).  One can show that the scalar field $\sigma$
decouples from the theory after gauge fixing terms are added upon
quantization.

The mass term for $V \equiv ( C_\mu,\, B_\mu,\, W^3_\mu )^T$, after
electroweak symmetry breaking $\langle \Phi \rangle = v/\sqrt{2}$, is
given by \cite{nath}
\begin{equation}
- \frac{1}{2} V^T  M  V  \equiv   -\frac{1}{2} 
  \left ( C_\mu,\, B_\mu,\, W^3_\mu \right ) \;
 \left( \begin{array}{ccc}
  M_1^2 & M_1 M_2 & 0 \\
 M_1 M_2 & M_2^2  + \frac{1}{4} g_Y^2 v^2 & -\frac{1}{4} g_2 g_Y v^2 \\
 0 & -\frac{1}{4} g_2 g_Y v^2  & \frac{1}{4} g_2^2 v^2 \end{array} \right ) \;
 \left( \begin{array}{c}
       C_\mu \\
       B_\mu \\
       W^3_\mu \end{array} \right ) \;.
\end{equation}
A similarity transformation can bring the mass matrix $M$ into a diagonal form
\begin{equation}
\left( \begin{array}{c}
                                        C_\mu \\
                                        B_\mu \\
                                        W^3_\mu \end{array} \right ) =
  O \, \left( \begin{array}{c}
                                        Z'_\mu \\
                                        Z_\mu \\
                                        A_\mu \end{array} \right ) 
\;\; , \;\;\;\;\;
 O^T  M \, O \, =
 {\rm diag}(m^2_{Z'},\, m^2_Z,\, 0 )  \; .
\end{equation}
The $m_{Z'}^2$ and $m^2_{Z}$ are given by
\begin{eqnarray}
m^2_{Z',\,Z} &=& \frac{1}{2} \biggr[ 
 M_1^2 + M_2^2 + \frac{1}{4} (g_Y^2 + g_2^2) v^2 \nonumber\\
&& \pm \sqrt{ (M_1^2 + M_2^2 + \frac{1}{4} g_Y^2 v^2 + \frac{1}{4} g_2^2 v^2)^2
 - ( M^2_1(g_Y^2 + g_2^2) v^2 + g_2^2 M_2^2 v^2 ) }  \biggr ]\;.
\label{zmass}
\end{eqnarray}
The orthogonal matrix $O$ is parameterized as 
\footnote
{We note that the middle column is chosen to be different 
from that of Ref.\cite{nath} by an overall minus sign.}
\begin{equation}
O = \left( \begin{array}{ccc}
              c_\psi c_\phi - s_\theta s_\phi s_\psi \;\; &  
              s_\psi c_\phi + s_\theta s_\phi c_\psi \;\; & 
            - c_\theta s_\phi \\
              c_\psi s_\phi + s_\theta c_\phi s_\psi \;\;& 
              s_\psi s_\phi - s_\theta c_\phi c_\psi \;\;& 
              c_\theta c_\phi \\
            - c_\theta s_\psi & c_\theta c_\psi & s_\theta \end{array}\right )
 \;,
\label{rotate}
\end{equation}
where $s_\phi =\sin\phi, c_\phi=\cos\phi$ and similarly for the angles
$\psi$ and $\theta$.  
The angles are related to the original parameters in the Lagrangian ${\cal L}_{\rm StSM}$ by
\begin{eqnarray}
\delta \equiv \tan \phi = \frac{M_2}{M_1} , && 
 \tan\theta = \frac{g_Y \cos\phi}{g_2}, \;\;
 \tan\psi = \frac{ \tan\theta\, \tan\phi\, m_W^2}
                 {\cos\theta [ m^2_{Z'} - m_W^2 (1+ \tan^2\theta) ]} \;,
                 \label{mixing-angles}
\end{eqnarray}
where $m_W = g_2 v/2$.  The Stueckelberg $Z'$ decouples from the 
SM when $\phi \to 0$, since
\[
 \tan\phi = \frac{M_2}{M_1} \to 0 \;\; \Rightarrow \;\;
 \tan\psi \to 0 \;\;\; {\rm and} \;\;\; \tan\theta \to \tan\theta_{\rm w} \;
\]
where $\theta_{\rm w}$ is the Weinberg angle.
 
The interactions of fermions with the neutral gauge bosons before rotating
to the mass eigenbasis are given by
\begin{equation}
- {\cal L}^{NC}_{\rm int} =
  g_2 W^3_\mu \, \bar \psi_f \gamma^\mu \frac{\tau^3}{2} \psi_f 
+ g_Y B_\mu \, \bar \psi_f \gamma^\mu \frac{Y}{2} \psi_f 
+ g_X C_\mu \, \bar \chi \gamma^\mu Q_X \chi  \;,
\end{equation}
where $f$ denotes the SM fermions.
The neutral gauge fields are rotated into the mass eigenbasis using 
Eq. (\ref{rotate}), and the above neutral current interaction becomes
\begin{eqnarray}
- {\cal L}^{NC}_{\rm int} &=& \bar \psi_f\, \gamma^\mu \left[ 
  \left( \epsilon_{Z'}^{f_{L}} P_L + \epsilon_{Z'}^{f_{R}} P_R\right)\, Z'_\mu
+ \left( \epsilon_{Z}^{f_{L}} P_L + \epsilon_{Z}^{f_{R}}P_R \right)\, Z_\mu
+ e Q_{\rm em} A_\mu \right ] \psi_f \nonumber\\
&+& 
 \bar \chi \gamma^\mu \left[ 
   \epsilon^\chi_\gamma A_\mu 
 +\epsilon^\chi_Z Z_\mu
 +\epsilon^\chi_{Z'} Z'_\mu \right ]\, \chi \; ,
\label{chiral1}
\end{eqnarray}
where
 \begin{eqnarray}
  \epsilon^\chi_\gamma &=& g_X Q^\chi_{X} ( - c_\theta s_\phi ) \; ,
 \nonumber \\
  \epsilon^\chi_Z &=& g_X Q^\chi_{X} ( s_\psi c_\phi + 
                s_\theta s_\phi c_\psi) \; , \nonumber \\
  \epsilon^\chi_{Z'} &=& g_X Q^\chi_{X} (c_\psi c_\phi - s_\theta s_\phi s_\psi
   ) \; , \nonumber \\
 \epsilon_Z^{f_{L,R}} &=& \frac{c_\psi}{\sqrt{ g_2^2 + g_Y^2 c_\phi^2}} \,
  \left( - c_\phi^2 g_Y^2 \frac{Y}{2} + g_2^2 \frac{\tau}{2} \right )
   + s_\psi s_\phi g_Y \frac{Y}{2} \; , \nonumber \\
 \epsilon_{Z'}^{f_{L,R}} &=& \frac{s_\psi}{\sqrt{ g_2^2 + g_Y^2 c_\phi^2}} \,
  \left(  c_\phi^2 g_Y^2 \frac{Y}{2} - g_2^2 \frac{\tau}{2} \right )
   + c_\psi s_\phi g_Y \frac{Y}{2} \;.
\label{chiral2}
\end{eqnarray}
In the above, we have used the relations
\[
 e = g_2 \, s_\theta = g_Y c_\phi c_\theta  \qquad {\rm and} \qquad
 Q_{\rm em} = 
\frac{\tau^3}{2} + \frac{Y}{2}  \;,
\]
where $Q_{\rm em}$ is the electric charge operator.
{}From Eqs.(\ref{chiral1})--(\ref{chiral2}), it is clear that in this 
class of model, the SM fermions 
interact with the hidden world through $Z'$ and the hidden fermion 
interacts with our visible world through $\gamma$ and $Z$. 
In our computation, we assume the following input parameters at the electroweak
scale \cite{pdg}
\[
  \alpha_{\rm em} ( m_Z) = \frac{1}{128.91},\;\; 
              G_F = 1.16637 \times 10^{-5} \,{\rm GeV}^{-2},\;\;
              m_Z = 91.1876\, {\rm GeV}\;, 
              \sin^2 \theta_{\rm w} = 0.231 \;,
\]
and the following three inputs related to the hidden sector
\[
  \delta \equiv \tan\phi,\;\; g_X, \;\; {\rm and} \;\; M_{Z'} \;.
\]
Since $Q^\chi_X$ always enters in the product form $g_X Q^\chi_X$, one can 
set $Q^\chi_X$ to be unity without loss of generality.
We derive from $\alpha_{\rm em}$, $G_F$, $m_Z$, and $\sin^2\theta_{\rm w}$
the values of 
\[
  e = \sqrt{4 \pi \alpha_{\rm em}}\;,\;
  v = \left( \sqrt{2} G_F \right)^{-1/2}\; , \;
  m_W = m_Z \sqrt{ 1 - \sin^2 \theta_{\rm w} }\;,\;\;
  {\rm and} \;\;\;
  g_2 = 2 m_W/v \;.
\]
We then fix the value of $g_Y$ by the following equation
\[
  e = \frac{g_2 g_Y c_\phi}{\sqrt{g_2^2 + g_Y^2 c_\phi^2}} \; .
\]
The other two angles $\theta$ and $\psi$ are determined from the last two 
formulas in Eq.(\ref{mixing-angles}).

It is clear from Eqs. (\ref{chiral1})-(\ref{chiral2}) that the chiral
couplings of the SM $Z$ boson are affected by the mixing.  In fact, even
the mass of the $Z$ boson is modified in this model, as shown by
Eq. (\ref{zmass}).  It has been shown in Ref. \cite{liu} that in order
to keep the $Z$ boson mass within the experimental uncertainty, the
mixing angle must satisfies
\begin{equation}
\label{delta-limit}
\delta \alt 0.061 \, \sqrt{ 1- (m_Z/M_1)^2 } \; .
\end{equation}
When $\delta$ is small and $m_{Z'}$ is large, $M_1 \approx m_{Z'} + O(g_2 v)$.
The constraint coming from the electroweak precision data is more or less
the same as in Eq. (\ref{delta-limit}) \cite{liu}. 

The limits obtained
in Ref. \cite{liu} also included the analysis from direct $Z'$ production 
at the Tevatron.  They showed that with the current Drell-Yan data,
\begin{eqnarray}
  m_{Z'} > 250 \;{\rm GeV} \;\; &{\rm for}& \;\; \delta \approx 0.035 \; ,
\nonumber \\
  m_{Z'} > 375 \;{\rm GeV} \;\; &{\rm for}& \;\; \delta \approx 0.06 \;.
\label{zprime-constraint}
\end{eqnarray} 
If including the presence of a hidden fermion that the Stueckelberg $Z'$ can
couple to, the limit from direct $Z'$ direction would be relaxed because
of the smaller production rate into visible lepton pairs \cite{liu}.  
In Sec. IV, 
we will show that with a hidden fermion $\chi$ fulfilling the dark matter
constraint, the $Z'$ would dominantly decay into the hidden sector fermion
pair provided that $m_{Z'} > 2 m_\chi$.  It would therefore entirely remove the
constraint in Eq. (\ref{zprime-constraint}) from direct production.  On the
other hand, if $m_{Z'} < 2 m_\chi$ the $Z'$ boson cannot decay into
the hidden sector fermions, and so the constraint in 
Eq. (\ref{zprime-constraint}) stands.

In the following numerical works, we will apply the constraints on
$\delta$ and $m_{Z'}$ given by Eqs. (\ref{delta-limit}) and
(\ref{zprime-constraint}) respectively, but when $m_{Z'} > 2 m_{\chi}$
the latter constraint will be dropped.

\section{Dark matter phenomenology}

\subsection{Milli-charged dark matter}
Milli-charged dark matter was first discussed by Goldberg and Hall
\cite{goldberg-hall}, motivated by the work of Holdom \cite{holdom} in
which milli-charged particles in the hidden sector can interact with
the visible photon due to kinetic mixing between the visible photon
and the hidden or shadow photon.  Numerous constraints for the
milli-charged particles, including accelerator experiments, invisible
decay in ortho-positronium, SLAC milli-charged particle search, Lamb
shift, big-bang nucleosynthesis, dark matter search, search of
fractional charged particles in cosmic rays, and other astrophysical
reactions, were summarized in
\cite{davidson} (see Fig. 1 of the first reference in \cite{davidson}).  
Study of the constraints on
milli-charged particles from neutron stars and CMB measurements 
were discussed in
Refs.\cite{Gould} and \cite{champs-cmb} respectively.
In summary, milli-charged particles of mass from
MeV to TeV with a fractional electric charge $(10^{-6}-10^{-1})$ of a
unit charge are still allowed.
We note that integral charged dark matter was contemplated in 
\cite{champs} and
composite dark matter was studied in \cite{khlopov}.
More recently, PVLAS \cite{PVLAS} reported a positive signal of vacuum
magnetic dichroism.  It has been suggested
\cite{gies-jaeckel-ringwald} that photon-initiated pair production of
milli-charged fermions with a mass range between 0.1 to a few eV and a
milli-charge $O(10^{-6}) $ of a unit charge can explain the signal.  However,
this signal has not been confirmed by other experiments like
the Q \& A experiment \cite{AhTou}. 
For detailed analysis of various experiments of vacuum magnetic dichroism, 
we refer our readers to Ref.\cite{ahlers-gies-jaeckel-ringwald}.

\subsection{Relic density and WMAP measurement}

The first set of processes we consider in our relic density
calculation are
\[
\chi \bar \chi \to Z', Z, \gamma \to f \bar f
\]
where $f$ is any SM fermion.
The amplitude for the annihilation $\chi (p_1)\; \bar \chi(p_2) \to
f(q_1)\; \bar f(q_2)$ can be written as
\begin{equation}
{\cal M} = \bar v(p_2)\, \gamma_\mu \,u(p_1) \times
      \bar u(q_1)\, \gamma^\mu \, \left ( \xi_L P_L + \xi_R P_R \right
              )\, v(q_2)
\end{equation}
where $P_{L,R} = (1 \mp \gamma_5)/2$, and
\begin{eqnarray}
  \xi_{L,R} &=& \frac{\epsilon_\gamma^\chi e Q_{\rm em}^f }{s}
           +\frac{\epsilon_Z^\chi \epsilon^{f_{L,R}}_Z }{s- m_Z^2}
           +\frac{\epsilon_{Z'}^\chi \epsilon^{f_{L,R}}_{Z'} }{s -
           m_{Z'}^2} \; .
\end{eqnarray}
The differential cross section is given by
\begin{eqnarray}
\frac{d \sigma}{d z} &=& \frac{N_f}{32\pi} \, \frac{\beta_f}{s
 \beta_\chi} \; \biggr [
 (\xi_L^2 + \xi_R^2 ) ( u_m^2 + t_m^2 + 2 m_\chi^2 ( s- 2 m_f^2 ) ) +
   4 \, m_f^2 \, \xi_L \xi_R ( s + 2 m_\chi^2 )
\biggr ]
\end{eqnarray}
where $\beta_{f,\chi} = ( 1 - 4 m_{f,\chi}^2/s )^{1/2}$, $N_f =
3\,(1)$ for $f$ being a quark (lepton), $t_m = t - m_\chi^2 - m_f^2 =
-s ( 1 - \beta_f \beta_\chi z)/2$, $u_m = u - m_\chi^2 - m_f^2 = -s (
1 + \beta_f \beta_\chi z)/2$, $s$ is the square of the center-of-mass
energy, and $z \equiv \cos\Theta$ with $\Theta$ the scattering angle.

We also consider pair annihilation of $\chi \bar \chi$ into two
neutral gauge bosons,
\begin{eqnarray}
\chi \bar\chi \to V_1 V_2 \;\;\; {\rm with} \;\;\; V_{1,2} = \gamma,
\; Z, \; Z'
\end{eqnarray}
in our relic density calculation 
\footnote{We have ignored the channel $\chi \bar \chi \to \gamma,Z,Z' \to W^+W^-$ which
may contribute to certain extent.}. 
The differential cross section is
given by
\begin{eqnarray}
\frac{d \sigma ( \chi \bar \chi \to V_1 V_2)}{d \Omega} &=&
\frac{S(\epsilon^\chi_{V_1})^2 (\epsilon^\chi_{V_2})^2 \beta_{V_1
     V_2}} {64 \pi^2 s \beta_\chi}\, \biggr \{
  -2 \left( 2 \, m_\chi^2 + m_{V_1}^2 \right) \left( 2 \, m_\chi^2 +
  m_{V_2}^2 \right) \left( \frac{1}{u_\chi^2} + \frac{1}{t_\chi^2}
  \right )
 \nonumber \\ &+& 2 \left( \frac{t_\chi}{u_\chi} +
\frac{u_\chi}{t_\chi} \right )
 - 4 \left( \frac{1}{u_\chi} + \frac{1}{t_\chi} \right ) \left ( 2 \,
   m_\chi^2 + m_{V_1}^2 + m_{V_2}^2 \right ) \nonumber \\
&-& \frac{4}{u_\chi t_\chi} \left( 2 \, m_\chi^2 + m_{V_1}^2 +
   m_{V_2}^2 \right) \left ( 2 \, m_\chi^2 - m_{V_1}^2 - m_{V_2}^2
   \right )
  \biggr \} \theta( 2 m_\chi - m_{V_1} - m_{V_2} ) \nonumber \\
  \label{chichi2VV}
\end{eqnarray}
where $\beta_{V_1V_2} = \lambda^{1/2}\left(1, m_{V_1}^2/s,\,
m_{V_2}^2/s \right)$ with 
$\lambda(a,b,c)=a^2+b^2+c^2-2(ab+bc+ca)$ is the Mandelstam function, 
$t_\chi$ and $ u_\chi$ are given by $t_\chi = t
- m_\chi^2$ and $u_\chi = u - m^2_\chi$ respectively, and $S$ is the
statistical factor.
We note that processes $\chi\bar \chi \to \gamma\gamma, ZZ$ are doubly
suppressed by the small mixing angles and $\chi\bar \chi \to Z'Z'$ are
either suppressed or forbidden by phase space, and therefore their
contributions are negligible in the annihilation rates.

The quantity that is relevant in the relic density calculation is the
thermal averaged cross section $\langle \sigma v \rangle$, where $v$
is the relative velocity of two annihilating particles.  In the
non-relativistic approximation, $v \simeq 2 \, \beta_\chi$.
To estimate the relic density of a weakly-interacting massive
particle, we use the following formula \cite{hooper}
\[
  \Omega_\chi h^2 \simeq \frac{0.1 \; {\rm pb}}{\langle \sigma v
  \rangle}\;.
\]
With the most recent WMAP \cite{wmap} result on dark matter density
\[
  \Omega_{\rm CDM} h^2 = 0.105 \pm 0.009 \;,
\]
where we have used the WMAP-data-only fit and taken $ \Omega_{\rm CDM}
= \Omega_{\rm matter} - \Omega_{\rm baryon}$, one can translate this
WMAP data into
\begin{equation}
  \langle \sigma v \rangle \simeq 0.95 \pm 0.08 \; {\rm pb} \;.
\end{equation}
In estimating the annihilation rate during the freeze-out, we assume
that the species has a velocity-squared $v^2 \simeq 0.1$.  To get a
crude estimation, we ignore the thermal average and evaluate
$\sigma v$ directly.

\begin{figure}[t!]
\centering
\includegraphics[width=3.2in]{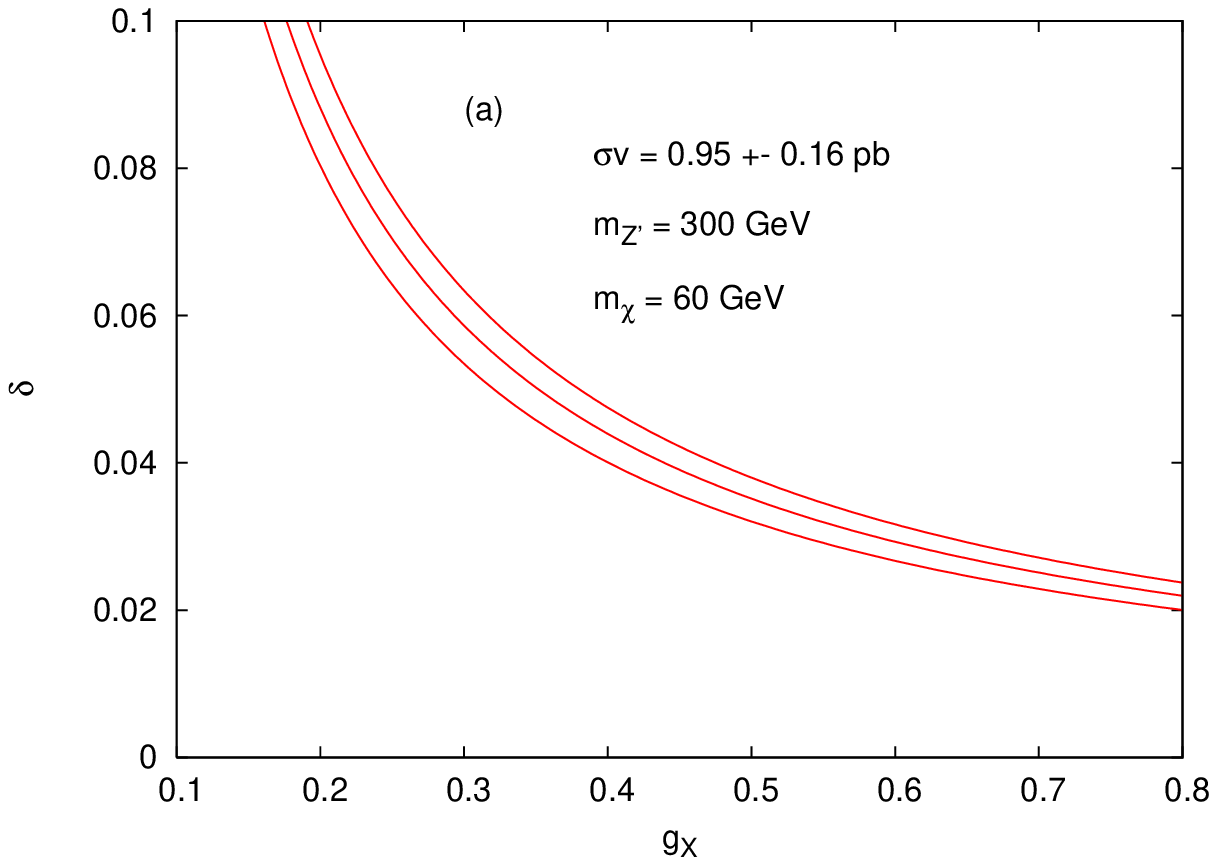}
\includegraphics[width=3.2in]{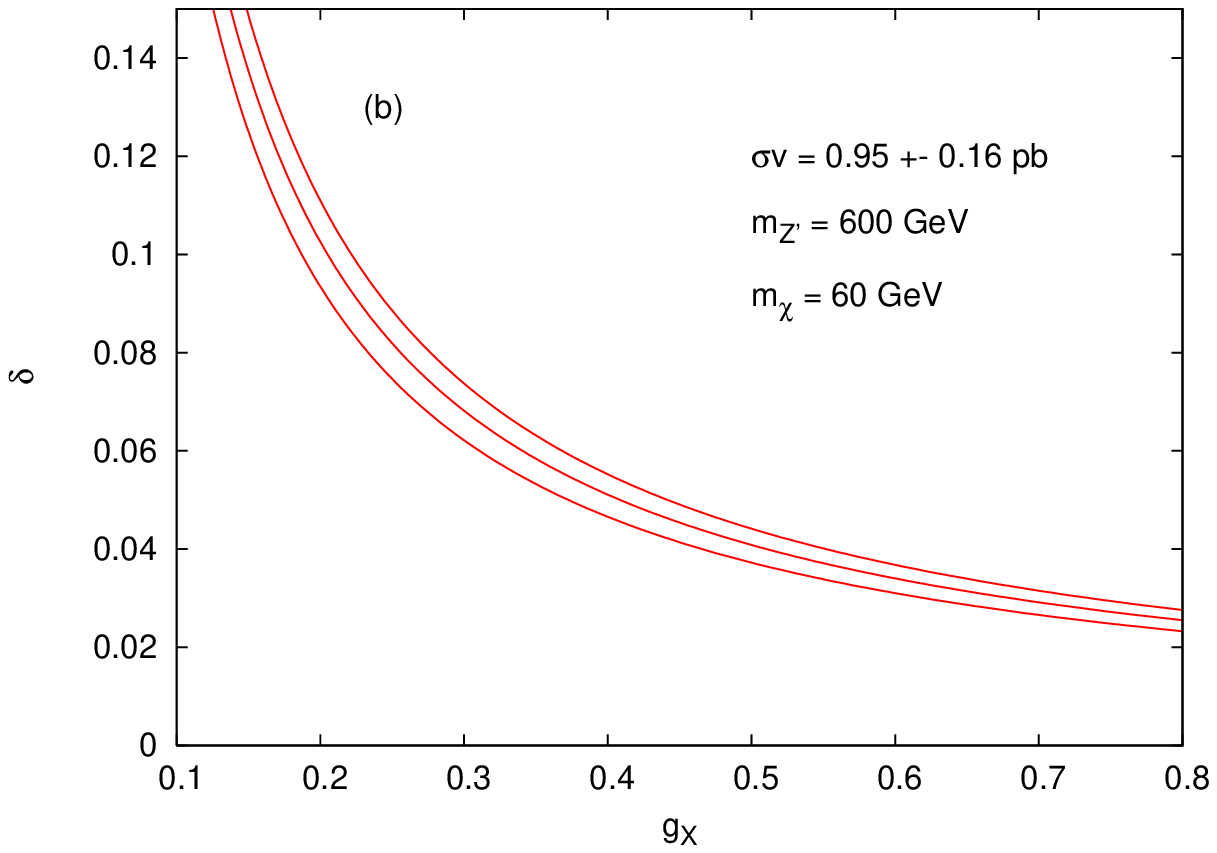}
\includegraphics[width=3.2in]{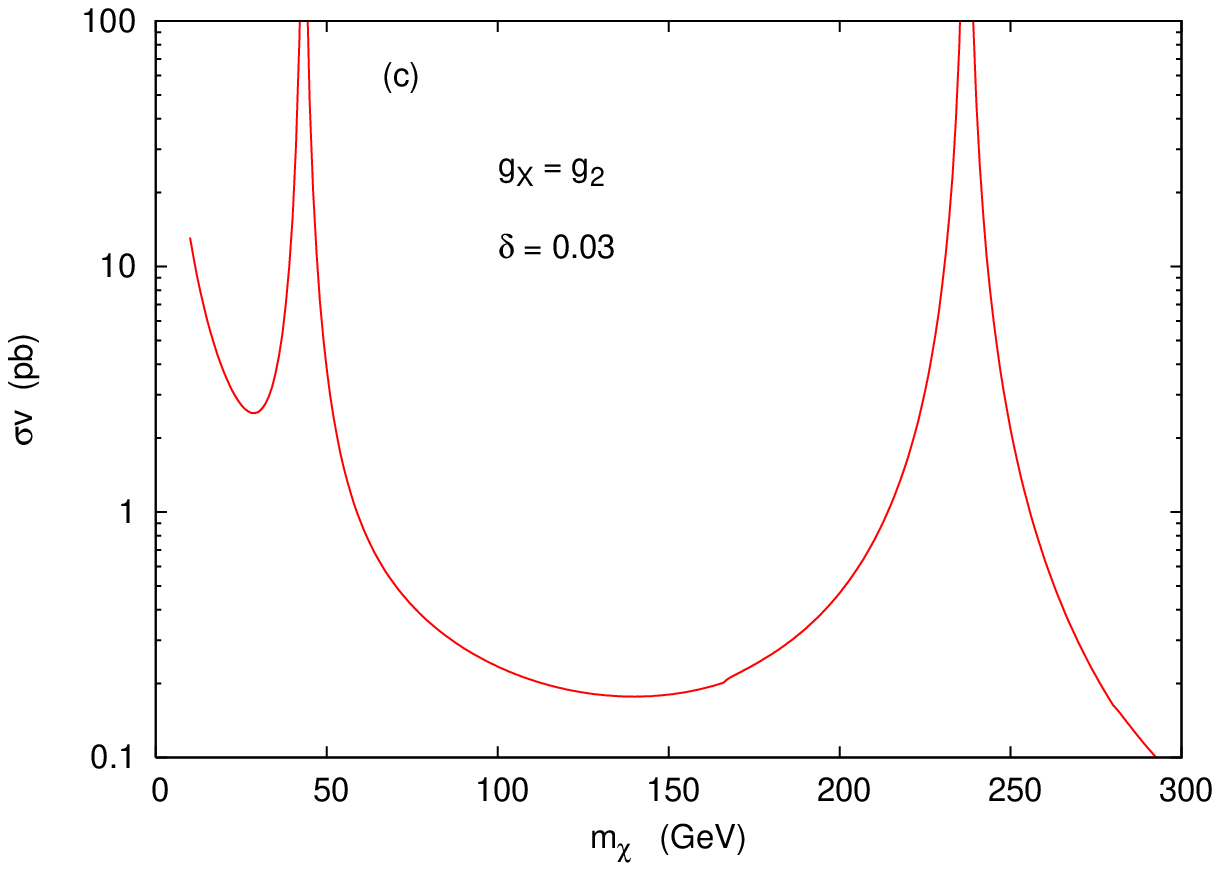}
\includegraphics[width=3.2in]{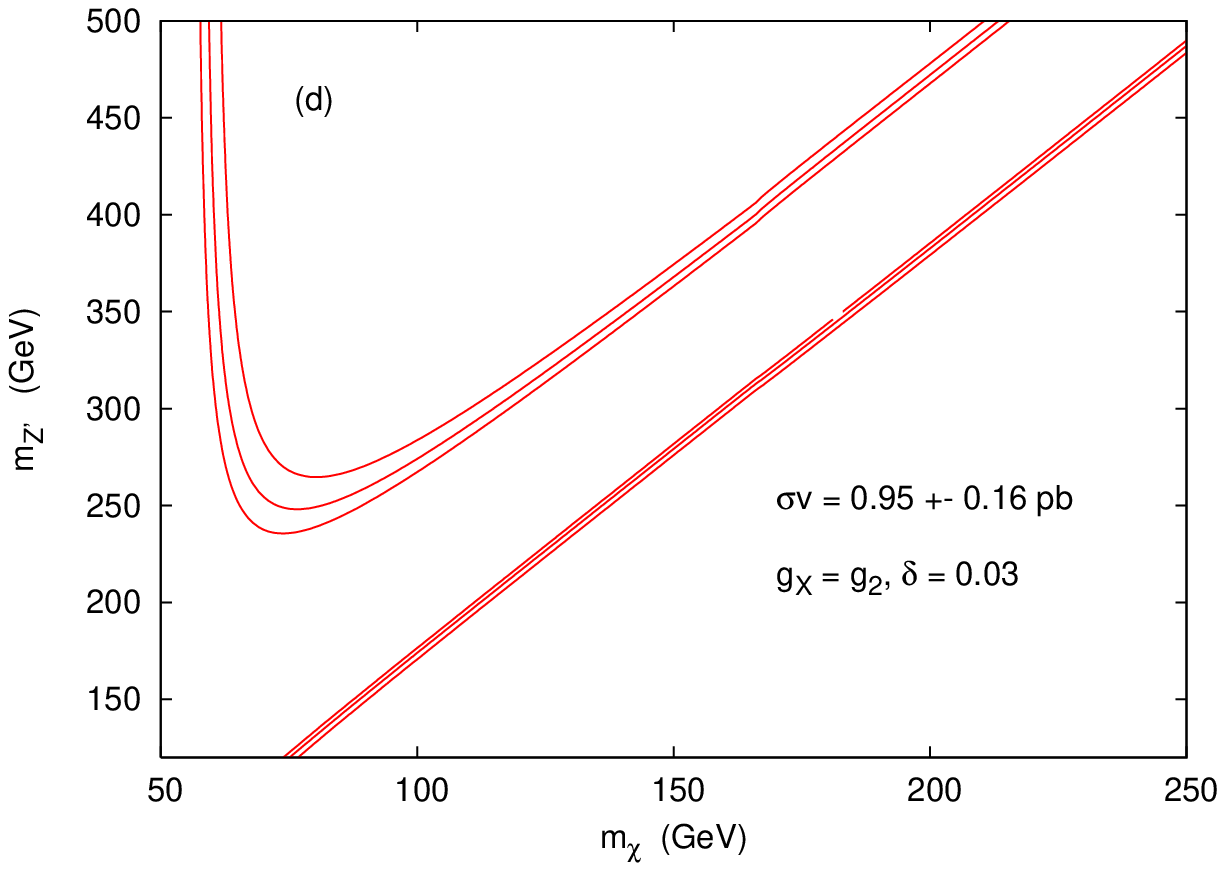}
\caption{\small \label{fig1} (a)--(b) are contours of $\sigma v$ =
$0.95 \pm 0.16$ pb (2 $\sigma$ range) in the plane of $(g_X,\,
\delta)$ for various $m_{Z'}$ and $m_{\chi}$. Part (c) shows the
annihilation rate $\sigma v$ versus $m_\chi$ with $m_{Z'} = 500$ GeV,
$g_X = g_2$, and $\delta = 0.03$.  Part (d) shows the contour of
$\sigma v$ = $0.95 \pm 0.16$ pb (2 $\sigma$ range) in the $(m_\chi,
m_{Z'})$ plane.  }
\end{figure}

In Fig.\ref{fig1}(a) and (b), we show the contours of $\sigma v = 0.95
\pm 0.16$ pb (2 $\sigma$ range) in the plane of $(g_X,\, \delta)$ for
various input values of $m_{Z'}$ and $m_{\chi}$.  We have included
$\chi \bar \chi \to \gamma Z', Z Z',$ and $f \bar f$, with $f=
\nu_e,\,\nu_\mu,\,\nu_\tau,\,e^-,\,\mu^-,\,\tau^-,\,u,\,d,\,s,\,c,\,b,$
and $t$ that are kinematically allowed.
From Fig.1(a) for $m_\chi=60$ GeV and $m_{Z'}=300$
GeV, we can see that $\delta \simeq 0.03$ and $g_X \simeq 0.6$ 
can give the correct amount of dark matter.
Similarly, from Fig.1(b) with the same $m_\chi = 60$ GeV and a larger $m_{Z'} = 600$ GeV, 
$\delta \simeq 0.03$ and a slightly larger $g_X \simeq 0.7$  can also do the job.  
For comparison, we note that $e\simeq 0.31$ and $g_2 \simeq 0.65$ in
the SM.  Thus, the value of the hidden $U_X(1)$ coupling $g_X$ that we
deduced from the WMAP measurement has the same order of the weak
coupling $g_2$ of the visible sector.
In Fig.\ref{fig1}(c), we show the annihilation rate $\sigma v$ versus
$m_\chi$ for $\delta =0.03$, $g_X = g_2$, and a fixed $m_{Z'} = 500$
GeV.  Clear resonance structures of $Z$ and $Z'$ are seen.
In Fig.\ref{fig1}(d), we show the contours of $\sigma v$ in the
$(m_\chi, m_{Z'})$ plane.
There are two branches: (i) the upper branch where $m_{\chi} <
m_{Z'}/2$ and the band relating $m_\chi$ and $m_{Z'}$ is relatively wide;
(ii) the lower branch where $2 m_\chi > m_{Z'}$ and the band relating 
$m_\chi$ and $m_{Z'}$ is quite narrow.
A narrow band implies the need of a fine-tuned relation between 
$m_\chi$ and $m_{Z'}$
in order to give the correct dark matter density.
In the latter branch, the Tevatron bound of 
$m_{Z'} > 250$ GeV for $\delta \approx 0.03$ has to be imposed.
Therefore, the case of $m_{\chi} < m_{Z'}/2$ is more preferred theoretically.

The hidden fermion $\chi$ couples to the photon via the mixing
angles $c_\theta s_\phi$, the value of which is about $0.9 \times
0.03 \approx 0.03$.  Therefore, effectively the fermion $\chi$
``acquires'' an electric charge of 
$g_X Q^\chi_{X} c_\theta s_\phi / e \approx 0.06$, when its coupling to the
photon is considered. 
Therefore, the range of $m_\chi \sim O(100)$ GeV and the 
size of effective electric charge
$\simeq 0.06$ implied by dark matter density requirement in our calculation 
are consistent with the constraints on milli-charged particles
\cite{davidson}.
\subsection{Indirect detection}

If milli-charged hidden fermions like $\chi$ and $\bar \chi$ are the
dark matter, 
their pair annihilation into 
$\gamma \gamma, \gamma Z$, and $\gamma Z'$ 
in regions of high dark matter density, e.g. the Galactic center,\
can give rise to monochromatic $\gamma$-ray line
that can reach our Earth for their indirect detection.
The cross sections for these processes can be obtained from 
Eq.(\ref{chichi2VV}) readily. 
The observed $\gamma$-ray flux along the line-of-sight between the Earth 
and the Galactic center is 
given by \cite{hooper}
\begin{eqnarray}
\Phi_\gamma (\psi, E) = \sigma v \frac{d N_\gamma}{d E_\gamma} 
 \frac{1}{4 \pi m_\chi^2} 
\int_{\rm line \; of \; sight} ds \rho^2(r(s,\psi)) \; ,
\label{22}
\end{eqnarray}
where the coordinate $s$ runs along the line of sight in a direction 
making an angle $\psi$ from the direction of the Galactic center, 
$dN_\gamma /dE_\gamma$ is the energy spectrum of the
$\gamma$-rays, and $v \approx  2 \beta_\chi$ is 
the relative velocity of the dark matter $\chi$ and $\bar \chi$,
and the present value of $v \approx 10^{-3}$.
The flux from a solid angle $\Delta \Omega$ is often expressed as
\begin{eqnarray}
\Phi_\gamma (\Delta \Omega, E) \approx 5.6 \times 10^{-12} 
\frac{d N_\gamma}{d E_\gamma}
\left( \frac{\sigma v} {{\rm pb}} \right)
\left( \frac{1 \, \rm TeV} {m_\chi} \right)^2 
{\overline J} (\Delta \Omega) \Delta\Omega \, {\rm cm^{-2} \, s ^{-1}} \; ,
\end{eqnarray}
with the quantity $J(\psi)$ defined by 
\begin{eqnarray}
J(\psi) = \frac{1}{8.5\, {\rm kpc}}
\left( \frac{1}{0.3 \, {\rm GeV/cm^3}} \right)^2
\int_{\rm line \; of \; sight} ds \rho^2(r(s,\psi)) \; .
\end{eqnarray}

\begin{figure}[t!]
\includegraphics[width=5in]{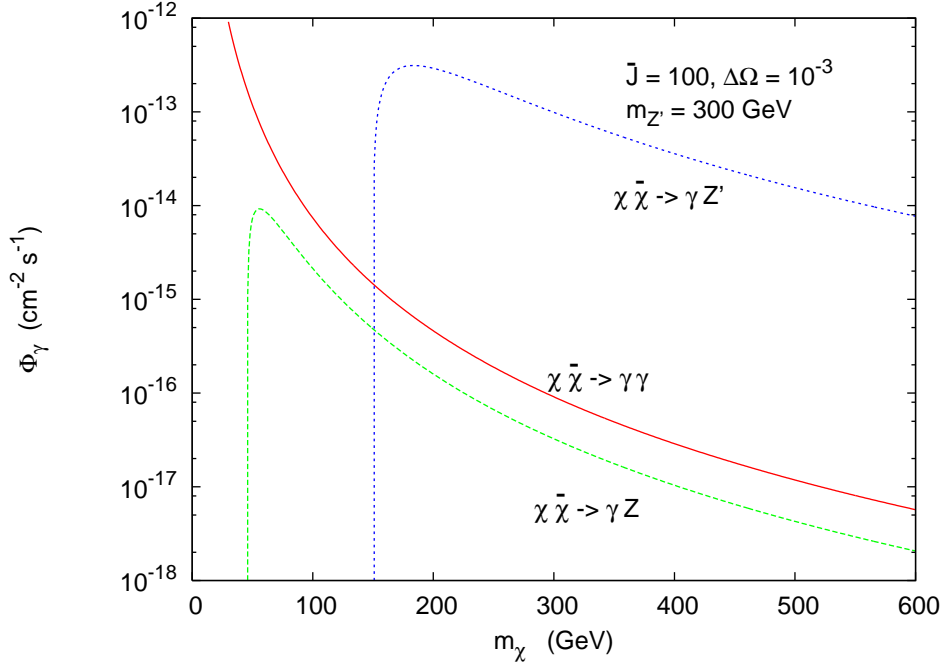}
\caption{\label{flux} \small
The resulting photon flux from annihilation processes $\chi \bar \chi \to
\gamma \gamma$, $\gamma Z$, and $\gamma Z'$.  We have used typical values
of ${\overline J} =100$, $\Delta \Omega = 10^{-3}$, and
the present value of $v \approx 10^{-3}$.}
\end{figure}

For the process of $\chi \bar \chi \to \gamma\gamma$, we would have a
mono-energetic $\gamma$-ray line with $dN_\gamma / dE_\gamma \approx 2 \delta(E_\gamma
- m_\chi)$.  Such a line, if observed, would be a distinctive signal
for dark matter annihilation.  Similarly, processes 
$\chi \bar \chi \to \gamma Z$ and 
$\chi \bar \chi \to \gamma Z'$ will have a photon energy spectrum as
$dN_\gamma / dE_\gamma  \approx 
\delta  (E_\gamma - m_\chi ( 1- m_{Z,Z'}^2/4 m_\chi^2) ) $. 
The contributions from these processes to the photon flux are shown 
in Fig. \ref{flux} with ${\overline J} =100$ and $\Delta \Omega = 10^{-3}$. 
In this plot, we have taken a moderate value for $\overline{J} = 100$
(averaged over $10^{-3}$ sr at the Galactic center).  
{}From Table 7 of Ref.\cite{hooper}, we know the value of $\overline{J}$ varies from
$2.166 \times 10$ (Kra profile) to $1.352 \times 10^3$ (NFW profile) and to $1.544 \times 10^{5}$ (Moore profile).  There are also cold dark matter profiles with dense spikes \cite{spikes} 
near the Galactic center due to the accretion by central black holes 
that can give rise significant enhancement to the quantity $\overline J$.  
With a rather conservative choice of $\overline{J} = 100$, the flux of the
$\gamma$-rays from the process $\chi \bar \chi \to \gamma\gamma$ 
is quite small due to double suppression of 
$(\epsilon^\chi_\gamma)^2$.  
The process $\chi \bar \chi \to \gamma Z$ contributes
at a somewhat lower flux level.
The process $\chi \bar \chi \to \gamma Z'$
can also contribute to the monochromatic photon flux, provided that 
$2 m_\chi > m_{Z'}$.  Since this process is only
suppressed by one power of the mixing angle, it could be
more substantial than the doubly-suppressed process
$\chi \bar \chi \to \gamma \gamma$. 
Note that since $2 m_\chi > m_{Z'}$, the Tevatron bound 
Eq.(\ref{zprime-constraint}) 
of $m_{Z'} >250$ GeV for $\delta \approx 0.035$  must be enforced. 
When kinematics allowed,
the photon flux from this process is three orders of magnitude 
higher than that from $\chi\bar \chi\to \gamma \gamma$.
For comparison, we note that the photon flux from the neutralino pair annihilation 
$\tilde \chi^0 \tilde \chi^0 \to \gamma \gamma$ \cite{bergstorm-ullio}
has been estimated to be 
about $1.5 \times 10^{-14}$ ($2 \times 10^{-13}$) 
${\rm cm}^{-2} {\rm s}^{-1}$ if the neutralino is a 
Higgsino-LSP (Wino-LSP), using the same moderate value of 
$\overline{J}=100$  \cite{Cheung:2005pv,Cheung:2005ba}. 

The expected sensitivities for the new Atmospheric Cerenkov Telescope (ACT)
experiments such as HESS \cite{hess} and VERITAS \cite{veritas1} are at the level of
$(10^{-14} - 10^{-13})$ cm$^{-2}$ s$^{-1}$ with an angular coverage of
about $10^{-3}$.  They are sensitive to the $\gamma$-rays
from a few hundred GeV to TeV.  On the other hand, the 
Gamma-ray Large Area Space Telescope (GLAST) experiment  \cite{glast} 
due for launch in Fall this year,
can probe $\gamma$-rays from 20 MeV to 300 GeV, 
but at a lower sensitivity level  about 
$2 \times 10^{-9} - 10^{-10} \;{\rm cm}^{-2} \, {\rm s}^{-1}$.
{}From Fig. \ref{flux}, there is a small
range of $m_\chi$ ( $m_\chi < 100 $ GeV) such that $\chi\bar \chi \to
\gamma\gamma$ contributes at a level larger than $10^{-14} \;{\rm
cm}^{-2}\, {\rm s}^{-1}$.  The process $\chi \bar \chi \to \gamma Z$
contributes at a level below the sensitivities of all these experiments for all ranges of
$m_\chi$, whereas the process $\chi \bar \chi \to \gamma Z'$ can
contribute at a much larger flux and it is indeed above the
sensitivity level of ACT experiments for $m_\chi < 600$ GeV.
Since GLAST can only be
sensitive to $\gamma$-rays of 300 GeV or less with lower
sensitivity, it is hard to detect the $\gamma$-rays from the lighter
milli-charged dark matter.
For heavier milli-charged dark matter,
the $\gamma$-rays can be above a few hundred GeV and thus above the
sensitivity reaches of HESS and VERITAS.

The continuum $\gamma$-ray background from astrophysical sources near Galactic
Center is largely an unsettled issue due to astrophysical uncertainties.  
There have been data showing
excess of $\gamma$-rays in different energy regimes near the Galactic center.  The EGRET
experiment \cite{egret} has reported an excess of $\gamma$-rays in the
region of the Galactic center, including the galactic longitude and latitude position
at $l=0^{\circ}$ and $b=0^{\circ}$ degrees.  The level
of excess is above the expectation of primary cosmic rays interacting
with interstellar medium.  The EGRET excess region is around 1 GeV.
However, there may be some other unknown sources of $\gamma$-rays around
the Galactic center.  It is hard to establish the fact that the excess is due to
dark matter annihilation, because the excess does not have specific
features. This is in contrast to the monochromatic $\gamma$-ray flux,
which is a clean signature of the dark matter annihilation.

In the Galactic center region, excesses of $\gamma$-rays were also
reported by VERITAS \cite{veritas2} in the range above 2.8 TeV and by
CANGAROO collaborations \cite{cangaroo} in the range of 250 GeV to 1
TeV.  Such excesses are also hard to be explained by conventional dark
matter candidates.  
The HESS Collaboration also had a measurement of TeV gamma rays from
the Galactic Center \cite{Aharonian:2004wa}, which is, to some extent,
in disagreement with the CANGAROO results.
It was pointed out \cite{Zaharijas:2006qb} that this TeV $\gamma-$ray excess is likely 
to be of astrophysical origin and thus it constitutes a background
for detecting dark matter annihilation. 
The origin of these backgrounds may be due to violent acceleration of cosmic
protons and other particles by the chaotic magnetic fields near the Galactic center black hole
 \cite{milky-way-black-hole}. 
After escaping the black hole environment and fly off into the interstellar medium, these extremely high energy protons collide with low energy protons (hydrogen gas) to form pions, which
subsequently decay into high energy $\gamma$-rays that can radiate in all directions.

Due to its unknown astrophysical origin, it is hard to establish accurately the true
continuum $\gamma$-ray background
which could be used for comparison with dark
matter annihilation.  Thus, using the continuum $\gamma$-ray signal is
difficult to provide strong evidence for dark matter,
unless the dark matter annihilation rate is very large.  
On the other hand, 
provided that the annihilation cross section is sufficiently large,
the monochromatic photon
line would be a ``smoking gun" signal for dark matter annihilation, 
since the energy of the $\gamma$-ray is uniquely determined by the mass of the 
milli-charged dark matter (and the $Z'$ mass as well in the 
$\chi \bar \chi \to \gamma Z'$ channel).  
Nevertheless, the EGRET and HESS continuum backgrounds still pose a 
serious challenge to detecting the monochromatic photon line due to
dark matter annihilation in the Galactic center region 
\cite{Zaharijas:2006qb}.  It was shown in Ref. \cite{Zaharijas:2006qb}
that in order for a photon line to be detected above the continuum 
background, the quantity 
$(\sigma v/10^{-29} \;{\rm cm}^3 \, {\rm s}^{-1} )\, \overline{J}  \,
\Delta \Omega$ must be larger than $10 - 100$.
This implies that the photon flux to be larger than
$1.9 \times ({\rm TeV}/m_\chi)^2 \times ( 10^{-14} - 10^{-13} )\;{\rm cm}^{-2}\;{\rm s}^{-1}$.  
From Fig. \ref{flux}, it is easy to see that for $m_\chi$ between 150 and 300 GeV,  
the photon flux in the $\chi \bar \chi \to \gamma Z'$ channel is within the
detectability level.

One may also give a rough estimate for continuum photon flux arises from 
the milli-charged dark matter annihilation into light quark pairs.
The continuum photon spectrum mainly comes from the light quark fragmentation into neutral pions, which subsequently decay into secondary photons.
The differential spectrum $d N_\gamma / d E_\gamma$ can be obtained by Monte Carlo event generators, e.g. P{\footnotesize YTHIA}, and then parameterized as a quark fragmentation function.
We can use Eq. (\ref{22}) with $dN_\gamma / dE_\gamma$ given by a fragmentation-like 
function \cite{fornengo}:
\begin{equation}
\frac{d N_\gamma} { d x} = \eta \, x^a \exp(b + c x + d x^2 + e x^3 ) \;,
\label{frag}
\end{equation}
where $x= E_\gamma/m_{\chi}$ and for a light quark, e.g. $u$ or $d$ 
quark at an energy of $500$ GeV,
 $\eta=1$, $a=-1.5$, $b=0.047$, $c=-8.7$, $d=9.14$, and $e=-10.3$.
These coefficients depend only mildly on the energy of the light quarks \cite{fornengo}.
Putting all these factors together, we estimate the integrated photon flux with 
$E_\gamma > 1$ GeV to be of the order of $10^{-10}$ $(10^{-11})$ cm$^{-2}$ s$^{-1}$ for 
$m_\chi = 100$ (500) GeV.  It is at most around or slightly below the sensitivity level of GLAST.  
Since the VERITAS and HESS experiments are sensitive to higher energy and the above 
spectrum Eq.(\ref{frag}) falls off rapidly as $x$ increases, 
their integrated photon fluxes are at least an order
of magnitude smaller than that of GLAST. 
Despite challenging by the uncertain astrophysical backgrounds, this continuous secondary photon spectrum together with the monochromatic photon line from milli-charged dark matter annihilation 
can be probed by the next generation of $\gamma$-ray experiments.

\section{Collider Phenomenology}

Phenomenology of the Stueckelberg $Z'$ with the presence of the hidden
fermion-antifermion $\chi$ and $\bar \chi$ that the $Z'$ can decay
into is quite different from the one studied before in
Refs.~\cite{nath,liu}.

\begin{figure}[t!]
\includegraphics[width=4in]{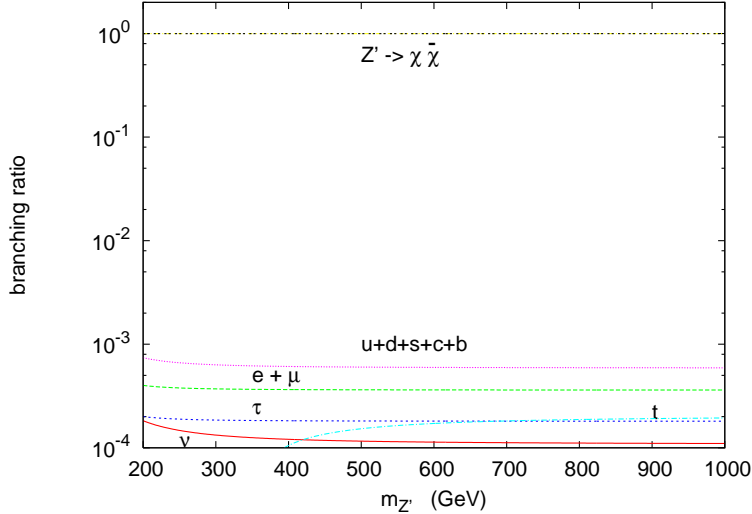}
\caption{\small \label{fig2}
Branching ratios for $Z'$ with $g_X= g_2$, $\delta = 0.03$, and 
$m_\chi = 60$ GeV. }
\end{figure}

The partial width of $Z'$ into a SM fermion pair $f\bar f$ is given by
\begin{equation}
\Gamma (Z' \to f \bar f)= \frac{N_f \beta_f}{24 \pi} m_{Z'} 
\biggr [ \left( {\epsilon_{Z'}^{f_L} }^2 + {\epsilon_{Z'}^{f_R} }^2 \right)
 \, \left( 1 - \frac{m_f^2}{m_{Z'}^2 } \right )
 + 6 \, \epsilon_{Z'}^{f_L}\, \epsilon_{Z'}^{f_R} \frac{m_f^2}{m_{Z'}^2} 
 \biggr ]
\label{zpw1}
\end{equation}
and into hidden fermion pair $\chi \bar\chi$ is simply
\begin{equation}
\Gamma (Z' \to \chi \bar \chi)= \frac{\beta_\chi}{12 \pi} m_{Z'} 
{\epsilon_{Z'}^{\chi}}^2 \, 
\left( 1 + \frac{2 \, m_\chi^2}{m_{Z'}^2 } \right ) \;.
\label{zpw2}
\end{equation}
Here, $\beta_{f,\chi} = (1 - 4 m_{f,\chi}^2/m_{Z'}^2)^{1/2}$.  The
total width of $Z'$ is evaluated by summing over all partial widths,
including both the SM modes and the hidden mode.  We show in
Fig. \ref{fig2} the various branching ratios for $Z'$ as a function of
its mass with the following inputs $g_X= g_2$, $\delta = 0.03$, and
$m_\chi = 60$ GeV.  Since the mixing angle is so small $(\delta =
0.03)$, the $Z'$ is mainly composed of the $C_\mu$ boson of the hidden
sector.  Hence, the $Z'$ dominantly decays into the hidden sector
fermion pair while it has only a small fraction of $10^{-3}$ into
visible fermions.  The strategy for the search of this $Z'$ would be
very different from all the previous conventional $Z'$ models
including the hidden Stueckelberg $Z'$ studied in \cite{nath,liu}.

Before we explore for the possible collider phenomenology of the 
Stueckelberg $Z'$ boson and the hidden sector fermion $\chi$, we have to
make sure that the new particles and the hidden sector interactions will
not upset the existing data.  

\subsection{Constraints from invisible decays of $Z$ and quarkonia} 

Firstly, the SM $Z$ boson that is observed at LEP would decay into a
pair of hidden fermions $\chi \bar \chi$, giving rise to additional
invisible width other than the neutrinos.  Because of the mixings
among the three neutral gauge bosons, the $Z$ boson can couple to the
$\chi \bar \chi$ pair via the mixing angle $s_\phi$.  We have
calculated the partial width of $Z \to \chi \bar \chi$ for $g_X =
g_2$, $\delta =0.03$ (consistent with the dark matter requirement),
and $m_\chi = 0 - 45$ GeV.  The partial width is about $0.24$ MeV,
which is much smaller than the uncertainty 1.5 MeV of the invisible
width of the $Z$ boson \cite{pdg}.  Even if we allow a larger mixing
angle $\delta = 0.061$ (its maximum value allowed by
Eq. (\ref{delta-limit})), the invisible width of $Z$ would be at most
1 MeV, which is still within the $1\sigma$ uncertainty of the data.
If the mass of $\chi$ is beyond half of the $Z$ boson mass, the
invisible width of the $Z$ boson would not constrain the model.

The hidden fermion $\chi$ can also couple to the photon via the mixing
angles $c_\theta s_\phi$, the maximum of which is about $0.9 \times
0.03 \approx 0.03$.  Therefore, effectively the fermion $\chi$
``acquires'' an electric charge of $g_X Q^\chi_{X} c_\theta s_\phi
/ e \approx 0.06 $ when its coupling to the photon is considered.  If
$\chi$ is very light, of the order of MeV, it could be produced in
$J/\psi$ and $\Upsilon$ decays as invisible particles.  Constraints on
invisible decays of $J/\psi$ and $\Upsilon$ exist (for a comprehensive
review on constraints on light dark matter: see Ref. \cite{Fayet}). A
very recent update on the $\Upsilon (1S)$ invisible width is given in
Ref. \cite{belle}.  The invisible widths of $J/\psi$ and $\Upsilon$
are respectively
\[
  B( J/\psi \to {\rm invisible} ) < 7 \times 10^{-3} \; \;\;\;
    {\rm and} \;\;\;\; 
  B( \Upsilon(1S) \to {\rm invisible} ) < 2.5 \times 10^{-3} \;.
\]
However, the partial width of $J/\psi$ into $\chi\bar\chi$ is
suppressed by the milli-charged factor of $(0.06)^2$ relative to the
partial width into $e^- e^+$.  Thus $B(J/\psi \to \chi \bar \chi)
\approx (0.06)^2 \times B(J/\psi \to e^- e^+) \approx 10^{-4}$, which
is well below the above limit.
The situation for $\Upsilon$ invisible decay is very similar: 
$B(\Upsilon (1S) \to \chi \bar \chi) \approx (0.06)^2 \times B(\Upsilon (1S)  \to
e^- e^+) \approx 10^{-4}$, which is also safe.
Indeed, a recent study \cite{bob} using 400 fb$^{-1}$ luminosity collected 
at the $\Upsilon(4S)$, the B-factory can limit
$B(\Upsilon(1S) \to {\rm invisible}) \lesssim 10^{-3}$.
There are also other decays modes, such as $J/\psi \;\; {\rm or} \;\;
\Upsilon \to \gamma + {\rm invisible}$, but it is straightforward to
check that with an effective charge of $0.06$ the experimental limits
of these radiative invisible modes do not constrain the model at all.
If the mass $m_\chi$ is above 5 GeV,
it has no constraint at all from these invisible decays of the quarkonia.

\subsection{Constraint from singly production of $Z'$ at LEPII}
\begin{figure}[t!]
\includegraphics[width=5in]{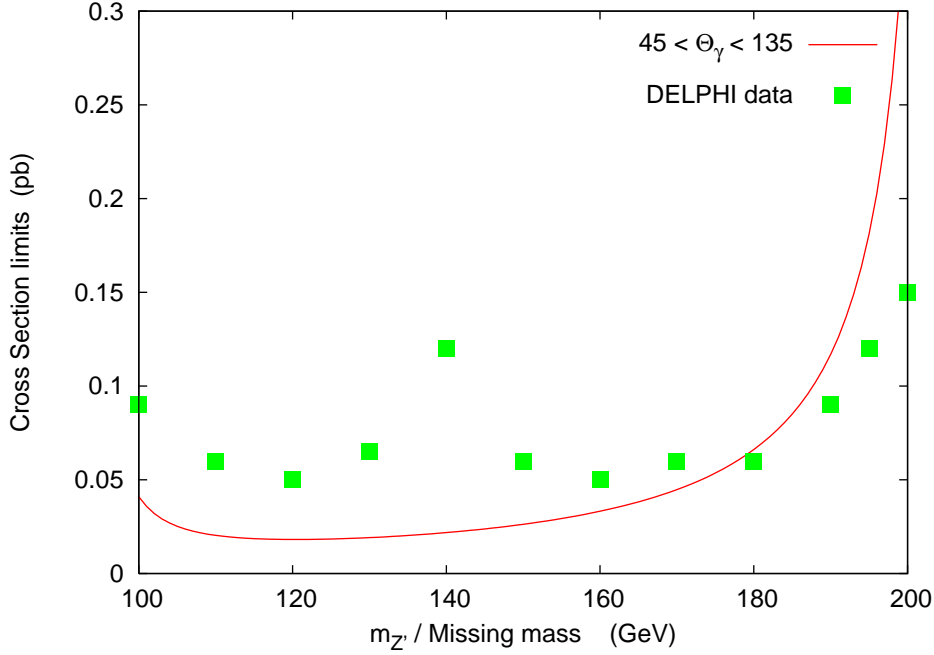}
\caption{\small \label{mp}
Comparison with the DELPHI data on the mono-photon production. The theory
prediction is for $g_X = g_2$ and $\delta = 0.03$.
}
\end{figure}

Singly production of the $Z'$ at LEPII is possible via $e^- e^+ \to \gamma Z'$
followed by the invisible decay of the $Z'$.  This process is very
similar to the SM process $e^- e^+ \to \gamma Z \to \gamma \nu \bar \nu$.  
The differential cross section for $e^- e^+ \to \gamma Z'$ is given by
\begin{equation}
 \frac{d\sigma( e^- e^+ \to \gamma Z')}{d \cos\Theta} =
\frac{\beta_{Z'} e^2 Q_e^2}{32 \pi s}\, \left( {\epsilon^{e_L}_{Z'} }^2
                                        + {\epsilon^{e_R}_{Z'} }^2 \right)
 \, \frac{1}{u t}\, \left[ t^2 + u^2 + 2 s m_{Z'}^2 \right] \;,
\label{eegz}
\end{equation}
where $\Theta$ is the scattering angle of the photon, 
$t =  - s  \beta_{Z'} (1 - \cos\Theta)/2$, 
$u = -  s \beta_{Z'} (1 + \cos\Theta)/2$, 
and $\beta_{Z'} = ( 1 - m_{Z'}^2 /s )$.
We show the production cross section at LEPII energy $\sqrt{s}=205$ GeV
in Fig. \ref{mp} as a function of $m_{Z'}$.  Since the $Z'$ would decay
into invisible $\chi\bar\chi$, the signal of which would be a mono-photon.
The recoil mass spectrum would then indicate the mass of the $Z'$. In
the figure, we also show the 95\% C.L. upper limits on mono-photon
production as a function of the missing mass obtained by DELPHI \cite{delphi}.
A small mass range of $Z'$, $180\;{\rm GeV} \alt m_{Z'} \alt 200$ GeV, is 
disfavored by the data.  However, one has to be cautious in this relatively
soft photon region where 
large theoretical uncertainties are expected to be important.

\subsection{Drell-Yan production of $Z'$ at the Tevatron}
\begin{figure}[t!]
\includegraphics[width=5in]{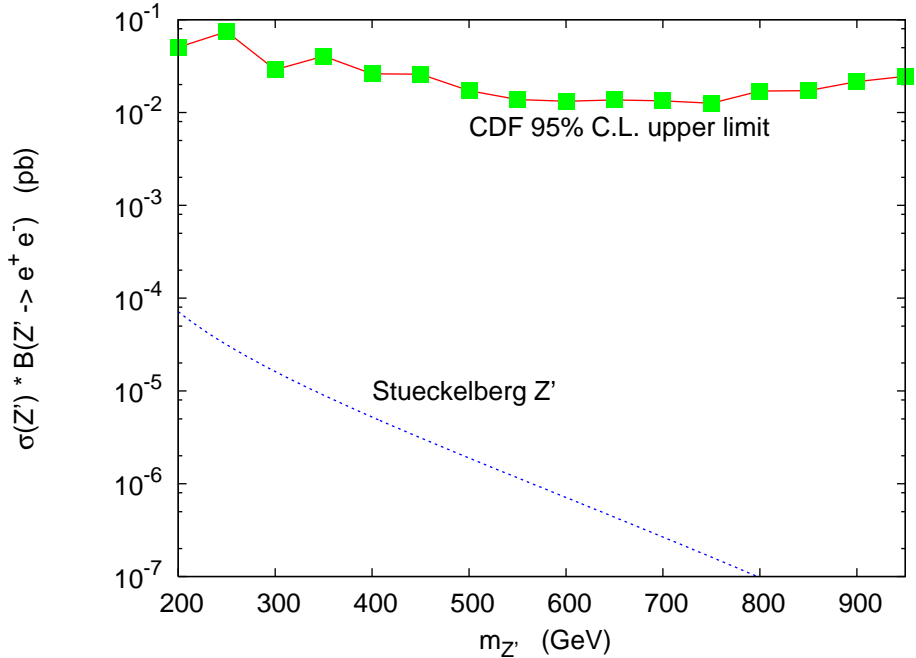}
\caption{\small \label{fig3}
Drell-Yan cross sections $p \bar p \to Z' \to e^- e^+$ versus $m_{Z'}$ for
$g_X = g_2$ and $\delta = 0.03$.
We also show the 95\% C.L. upper limits on $\sigma(Z')\cdot B(Z' \to 
e^- e^+)$ from the CDF preliminary results \cite{cdf-z}.
}
\end{figure}

The production cross section of $Z'$ followed by the leptonic decay at 
the Tevatron is given by
\begin{equation}
\sigma(p\bar p \to Z' \to \ell^- \ell^+) =
  \frac{1}{144} \frac{1}{s} \frac{m_{Z'}}{\Gamma_{Z'}} 
 \left( {\epsilon_{Z'}^{\ell_L}}^2 + {\epsilon_{Z'}^{\ell_R}}^2 \right )
\sum_{q=u,d,s,c}
 \left( {\epsilon_{Z'}^{q_L}}^2 + {\epsilon_{Z'}^{q_R}}^2 \right )
 \int^1_{r}\, \frac{dx}{x}  f_{q} (x) \, 
     f_{\bar q} \left( \frac{r}{x} \right )
\end{equation}
where $\sqrt{s}=1960$ GeV, $r = m_{Z'}^2/s$, $\Gamma_{Z'}$ is the total
width of $Z'$ given in Eqs. (\ref{zpw1}) and (\ref{zpw2}),
and $\epsilon_{Z'}^{f_{L,R}}$ can be found in Eq. (\ref{chiral2}).
This Drell-Yan cross section for the $Z'$ boson is plotted in
Fig. \ref{fig3}, where the 95\% C.L. upper limits on $\sigma(Z')\cdot
B(Z' \to e^- e^+)$ from the CDF preliminary results \cite{cdf-z} are
also shown.  It is clear that the present CDF limits do not constrain
the model at all.  This is in sharp contrast to the results studied in
Ref. \cite{liu} because the $Z'$ boson that we consider here has a
very small branching fraction into charged lepton pair.  The $Z'$
boson would decay preferably into the hidden sector fermions instead
of visible particles.  On the other hand, the Stueckelberg $Z'$ in
Ref. \cite{liu} only decays into the SM particles.  In our case the
$Z'$ only has a branching ratio of $\sim 10^{-4} - 10^{-3}$ into
leptonic pairs, and it would not be easily detected in the Drell-Yan
channel.  Neither the hadronic decay modes of $Z'$ can afford it to be
detected.
Even in the future runs of the Tevatron with a sensitivity 
reaching the level of  $10^{-3} - 10^{-2}$ pb, it is still not possible to
detect this kind of $Z'$ boson through the Drell-Yan channel. 

\subsection{Singly production of $Z'$ at LHC and ILC}

\begin{figure}[t!]
\includegraphics[width=5in]{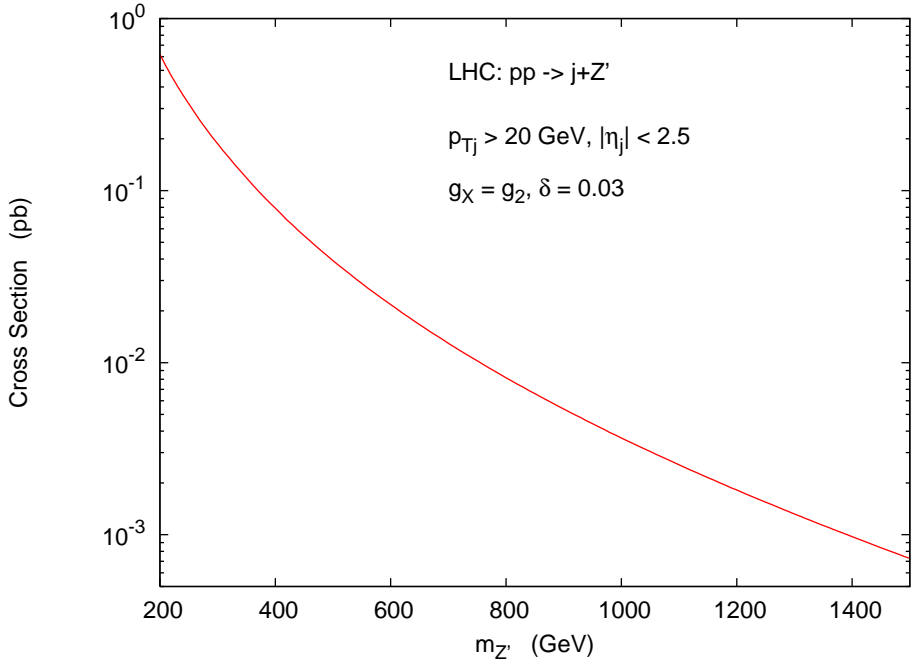}
\caption{\small \label{monojet}
Production cross section for the process $pp \to j + Z'$ followed
by invisible decay of the $Z'$ with $g_X = g_2$ and $\delta = 0.03$ at the 
LHC.
We imposed $p_{Tj} > 20$ GeV and $|\eta_j| < 2.5$ on the jet.}
\end{figure}

Perhaps one has to rely on the invisible decay mode of the
Stueckelberg $Z'$ of this model to identify its presence.  Here we
calculate the predictions of singly $Z'$ production at the LHC and
ILC.  Other than the Drell-Yan process that we have considered, the
next relevant process to probe this invisible $Z'$ is via $q\bar q \to
g Z'$ followed by $Z' \to \chi\bar\chi$, which gives rise to monojet
events.  The subprocess cross section can be easily adapted from
Eq. (\ref{eegz}):
\begin{equation}
\frac{d\hat \sigma( q \bar q \to g Z')}{d \cos\theta^*} =
\frac{\beta_{Z'} g_s^2} {72 \pi \hat s}\, \left( {\epsilon^{q_L}_{Z'} }^2
                                        + {\epsilon^{q_R}_{Z'} }^2 \right)
 \, \frac{1}{\hat u \, \hat t}\, \left[ \hat t^2 + \hat u^2 + 2 \, \hat s 
\, m_{Z'}^2 \right] \;.
\label{qqbar2gzp}
\end{equation}
Other cross channels, e.g., $q g \to q Z'$, can be obtained from 
Eq. (\ref{qqbar2gzp}) using the
crossing symmetry. 
The branching ratio $B(Z' \to \chi \bar \chi)$ is very close to unity.  We 
show in Fig. \ref{monojet} 
the production rate of monojet events versus $m_{Z'}$ with 
$g_X  = g_2$ and $\delta = 0.03$ at the LHC under the jet cuts of
$p_{Tj} > 20$ GeV and $|\eta_j| < 2.5$.  The $q\bar qZ'$ coupling is
suppressed by the small mixing angle, the same as in the Drell-Yan process,
but unlike the Drell-Yan process, this monojet amplitude is suppressed
by only one power of the mixing angle instead of two.  Therefore, the rate
is not negligible.  Also, the true SM background for monojet is rather
rare.  Thus, the monojet event actually signals the presence of such
an invisible $Z'$.

\begin{figure}[t!]
\includegraphics[width=5in]{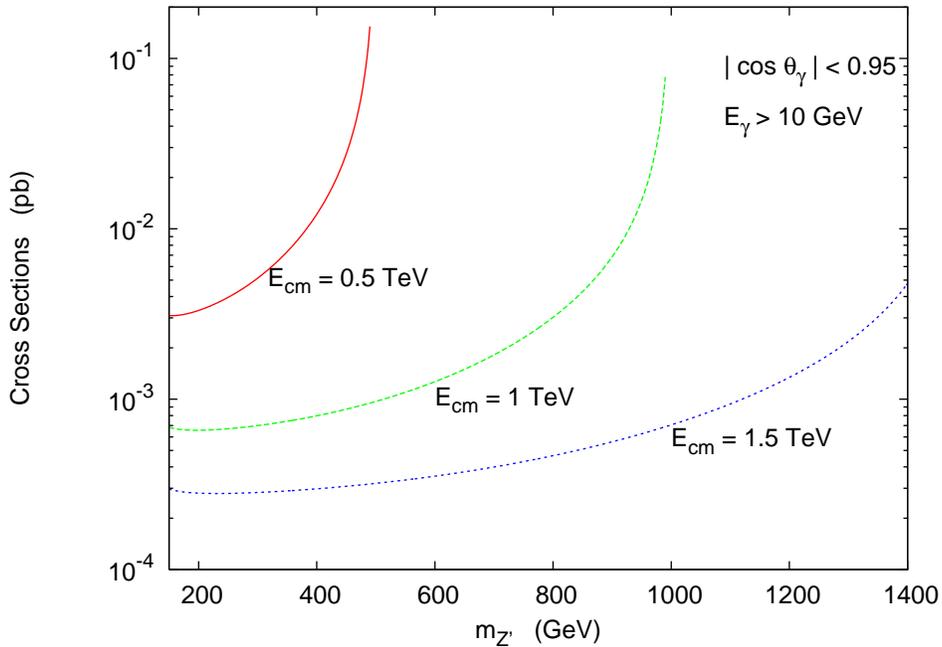}
\caption{\small \label{monophoton}
Production cross section for the process $e^- e^+ \to \gamma + Z'$ followed
by invisible decay of the $Z'$ with $g_X = g_2$ and $\delta = 0.03$
at ILC ($\sqrt{s} = 0.5,\, 1,\, 1.5$ TeV).
We imposed $E_{\gamma} > 10$ GeV and $|\cos_\gamma| < 0.95$ on the photon.}
\end{figure}

Another place to detect such an invisible $Z'$ is at the ILC with the
process $e^- e^+ \to \gamma Z' \to \gamma \chi \bar \chi$,
which we have considered above for the mono-photon limits from LEP. 
We extend the energy to $0.5 -1.5$ TeV and calculate the event rates for
the mono-photon final state.  We show in Fig. \ref{monophoton} 
the production rates
at $\sqrt{s} = 0.5, 1, 1.5 $ TeV with $g_X = g_2$ and $\delta = 0.03$.  

\section{Conclusions}

We have proposed an extension of the Stueckelberg $Z'$ standard model
by adding a pair of fermion and antifermion in the hidden sector,
which has only a $U(1)_X$ symmetry. The stability of the hidden
fermion pair with its weak sized interaction makes it a suitable dark
matter candidate with a correct amount of dark matter density.  We
have calculated the photon flux from the Galactic center due to the
annihilation of this milli-charged dark matter.  If $2 m_{\chi} <
m_{Z'}$, there is only a small range of $m_\chi$ that the photon flux
is above the sensitivity level of the future $\gamma$-ray experiments.
However, if $2 m_{\chi} > m_{Z'}$ there is a wide range of $m_\chi$
that the photon flux is above the sensitivity level.  The collider
phenomenology may be different from those studied in Ref. \cite{liu},
because the dominant decay of the $Z'$ is into the invisible $\chi\bar
\chi$ if kinematically allowed.  In this case, the present Drell-Yan
data cannot constrain the model at all.  We have proposed the monojet
signal at the LHC and the mono-photon signal at the future ILC to
probe this invisibly decaying Stueckelberg $Z'$ boson.

We close with some comments.
\begin{enumerate}
\item 
Since only a $U_X(1)$ symmetry is assumed in the hidden sector, each hidden 
fermion is stable against decay.  Therefore, if we assume more  
fermion pairs in the hidden sector, their relic densities are additive.
Thus, a larger coupling constant
is needed to ensure larger annihilation cross sections. 
One can also consider multiple hidden Stueckelberg $U(1)$ extension of the SM. 
We refer to Ref.\cite{nath} for the discussion for this possibility.

\item
When $m_{Z'} < 2 m_{\chi}$, the $Z'$ decays dominantly into visible
particles.  It can be easily detected in the Drell-Yan channel.  The
existing data constrains the model, as given by
Eq. (\ref{zprime-constraint}) originally obtained by the authors in
Ref.\cite{liu}. Photon flux from pair annihilation of $\chi \bar \chi
\to \gamma Z'$ at the Galactic center is also within reach at the next
generation of $\gamma$-ray experiments.

\item
When $m_{Z'} > 2 m_{\chi}$, the $Z'$ decays dominantly into invisible
$\chi\bar \chi$.  The present Drell-Yan data cannot constrain the model,
neither can the invisible decays of $J/\psi$ and $\Upsilon$ for a very light $\chi$. 
However, the mono-photon production limits obtained by DELPHI disfavors a small
range of $180\;{\rm GeV} \alt m_{Z'} \alt 200$ GeV.
We anticipate that in the future ILC the missing mass spectrum can efficiently
constrain this type of invisibly decaying $Z'$ boson.

\item
The hidden fermion appears to have a milli-charge as it acquires a small
effective coupling to the photon through the mixing induced by the combined 
Higgs and Stueckelberg mechanisms.  With a mass of $O(100)$ GeV and an 
effective charge $0.06$ of a unit charge, the hidden fermions are consistent
with the existing constraints on milli-charged particles \cite{davidson}.
As milli-charged particle is of very recent interests 
\cite{gies-jaeckel-ringwald}, an update on the terrestrial and astrophysical 
constraints on this hidden milli-charged particle is desirable.

\end{enumerate}

\newpage

\section*{Acknowledgment}
We would like to thank Pran Nath for encouragement and 
Holger Gies for useful comments on the manuscript.
K. C. also thanks K. S. Cheng and the Centre of Theoretical and Computational
Physics at the University of Hong Kong for hospitality.
This research was supported in part by the National Science Council of Taiwan
R.~O.~C.\ under Grant No.\ NSC 95-2112-M-007-001 
and by the National Center for Theoretical Sciences.

{\it Note added.} Stueckelberg $Z'$ extension with kinetic mixing has been studied recently in 
\cite{feldman-liu-nath}. Wherever overlaps in the parameter space, 
the authors in \cite{feldman-liu-nath} found agreements with the analysis presented in our work.


\end{document}